# Expanded Distance-based Phylogenetic Analyses of Fossil *Homo* Skull Shape Evolution

Peter J. Waddell[1]

pwaddell.new@gmail.com

[1]Ronin Institute, 1657 Upland Dr, Columbia, SC 29 204

Analyses of a set of 47 fossil and 4 modern skulls using phylogenetic geometric morphometric methods corroborate and refine earlier results. These include evidence that the African Iwo Eleru skull, only about 12,000 years old, indeed represents a new species of near human. In contrast, the earliest known anatomically modern human skull, Qafzeh 9, the skull of "Eve" from Israel/Palestine, is validated as fully modern in form. Analyses clearly show evidence of archaic introgression into Gravettian, pre Gravettian, Qafzeh, and Upper Cave (China) populations of near modern humans, and in about that order of increasing archaic content. The enigmatic Saldahna (Elandsfontein) skull emerges as a probable first representative of that lineage, which exclusive of Neanderthals that, eventually lead to modern humans. There is also evidence that the poorly dated Kabwe (Broken Hill) skull represents a much earlier distinct lineage. The clarity of the results bode well for quantitative statistical phylogenetic methods making significant inroads in the stalemates of paleoanthropology.

# 1 Introduction

In previous work novel phylogenetic methods for analyzing geometric morphometric data were developed (Waddell 2014). They were applied to 13 skulls representing the breadth of diversity in the genus *Homo*. These pairwise distance-based methods include some that fall within the definition of composite maximum-likelihood methods (Larribe and Fearnhead 2011, Varin et al. 2011), which are a reliable class of methods for making parameter estimates under models which might otherwise be too complex to apply. One newly developed approach was the use of likelihood to evaluate how simple monotonic transformations of the data might improve evolutionary tree estimation (Waddell 2014). The results of these analyses included considering the position of an enigmatic skull from Africa that is only around 12,000 years old (Brothwell and Shaw 1971, Shaw and Daniels 1984, Havarti et al. 2011, 2013). This Iwo Eleru skull has been shown to almost represent a distinct species from modern humans using phylogenetic methods applied to the data (Waddell 2014). This analysis extends from a subset of 11 fossil skulls to 47 to allow further testing to this exciting new proposal.

In this article a range of other interesting hypotheses are evaluated. One of these is a more thorough investigation of the affinities of the first known specimen of an anatomically modern human, that is the species *Homo sapiens*. On this topic, the work of Schwartz and Tattersall (2002, 2003, 2005) brings together the points essential for reliably identifying morphological modernity. Through the most extensive and detailed examination of *Homo* fossils yet undertaken, then show that the Qafzeh 9 specimen, about 90,000 years old, found in Israel/Palestine and usually designated female, is the first fairly intact skull and mandible that can said to conform to the expectations of a modern human form. Other less perfectly preserved specimens from the same site cannot be dismissed as modern human in form, but some, such as Qafzeh 6, can be. While Schwartz and Tattersall (2003) identify multiple morphs at this site, the analyses of Waddell (2014) give the first clear quantitative results based, on an evolutionary model, and these show that Qafzeh 6 is quite probably a hybrid. In this case it seems to be a hybrid of ~2/3 near anatomically modern human and 1/3 near Neanderthal archaic human.

Including further information on the forms and history of the Qafzeh and Skhul samples (Schwartz and Tattersall 2003), the hypothesis to reject emerges as it being a hybrid of a population derived from anatomically modern humans interbreeding with Middle Eastern Neanderthals. This is not the same as saying that these are modern humans interacting with Neanderthals. That is because what would be identified as anatomically modern human almost certainly predates the emergence of modern humans (the crown group *Homo sapiens*), that is, following the same lines of argument used to derive the modern super-ordinal classification of placental mammals (Waddell et al. 1999, 2001). That this is probably the case here is further attested to by no signs of modern human behavior, which becomes common later all around the world in association with populations showing clear signs of anatomical modernity (Bar-Yosef and Belfer-Cohen 2013). In this sense, Qafzeh 9 might be dubbed the skull of "Eve," the nearest thing we have yet to a mother of anatomically modern humans. Somewhat symbolically, this skull comes from the land of Israel/Palestine at the centre of the cultures originating the "Adam and Eve" creation myth, but about 80,000 years earlier than that "world" was created.

Other skulls of particular interest are how the loosely knit *Homo heidelbergensis*-like forms associate in phylogenetic analyses. There are five such well known specimens included here, the Kabwe (Broken Hill), Dali, Petralona, SH5 and Saldahna skulls (LH 18, sometimes associated with this form is not generally referred to as such here since the analyses of Waddell (2014) give the first solid quantitive evidence based on quantitative evolutionary models that it is not, while authors such as Schwartz and Tattersall (2004) give a widely informed qualitative assessment of why they consider it falls with an African lineage eventually leading to modern humans. The Saldahna (Elandsfontein) skull is sometimes considered a likely female of the form represented by Kabwe, so it will be interesting to see if any evidence of this emerges here.

The data at hand include a good sampling of early anatomically and near anatomically modern humans in different areas allowing assessment of potential archaic contribution (e.g., Wolpoff 1999). These include very early modern Europeans (pre-Gravettian cultures), early Europeans (Gravettian cultures commonly known as Cro Magnons), unusual Chinese forms (the Zhoukoudian Upper Cave skulls, which show a mixture of modern and archaic features), plus the already mentioned Qafzeh/Skhul forms (Klein 2009). Trinkaus (2007) argues persuasively that the European forms show features that could have been derived from Neanderthals, while some recent sequence data suggests that in at least one case, this is confirmed (on an Oase specimen, Fu et al. 2015). Here, the extent to which archaic interbreeding might have altered the general form of the skulls from these populations is further assessed. Also assessed for the first time with a quantitative evolutionary model will be the relative extent of archaic genes in these four distinct populations, and a discussion of what might have happened later, potential via the agencies of natural selection.

The present data set is an excellent one on which to further assess patterns of modern to archaic interbreeding. Trinkaus (2007) notes, for the pre-Gravettian samples, that Oase 2, Cioclovina 1 and Muierii 1 all show long flattened frontal bones, while occipital bunning is present in Muierii 1 and Mladec 5, with a reduced pronouncement (hemi-bunning) in Cioclovina, Mladec 1, and Oase. The present analyses do not seek nor weight these features directly, so localized bunning, for example, is expected to introduce a relatively small disturbance in the overall Procrustes distance. Thus, whether the genes involved in these features explain much of the variance in overall shape, for example, towards the more jellybean shaped heads of Neanderthals, can be tested herein. Since the shape of the frontal would be expected to influence the current Procrustes distances most, Muierii might be predicted to be the deepest diverging as it is said to show clearest evidence of both long flattened frontals and bunning, with Oase and Cioclovina being second and then Mladec 5 and Mladec 1.

## 2  Materials and Methods

The data used here are a set of 3D morphometric measurements based on 19 landmarks and 78 semi-landmarks from a set of modern and fossil skulls of the genus *Homo* (Harvati et al. 2011). The data largely follow 3D points on the curves of the major sutures and spatial bone boundaries on the top of the skull, these being the mid-line (sagittal suture and continuation into the occipital to the inion), across the top of the brow (supraorbital ridge), the suture that runs from the near the top of the skull down the sides towards the top of the temple region and marks the boundary of the frontal and the parietal bones (coronal suture) and finally, the boundary of the parietal bones and the occipital squama at the rear of the skull (lambdoid suture). From the 3D coordinates, pairwise Procrustes distances were calculated using a variety of software, which agreed closely with the numbers from Harvati et al. (2011). Procrustes distances are the minimum possible unweighted least squares distance between the corresponding landmarks and semi landmarks of one skull to the other achieved by rescaling, translating and rotating. In this particular case, various transformations of the sum of squares distance were made and assessed. These are available from the author upon request.

A number of programs and scripts were used for these distance analyses. These include PAUP*, which is now at alpha version 146, a major prerelease of the next version (Swofford 2002). It incorporates a range of flexibly weighted least squares models (Waddell, Azad and Khan 2011). In addition, a variety of Perl scripts were used, which enabled routines such as residual resampling and output of a trees design matrix, from Waddell, Tan and Ramos (2012) and Waddell and Tan (2013). Some specific calculations were also made in Apache Open Office.

## 3 Results

The data set used here are the pairwise Procrustes distances of 51 skulls from the genus *Homo*. All but four of these are fossil (or sub-fossil) skulls all dated to over 10,000 years old (e.g.

Schwartz and Tattersall 2002, 2003, 2005, Millard 2008). The four modern skulls are the same male and female from a Khoisan sample used in Waddell (2014) plus a recent Australian aboriginal male (labeled in the data "AboM012H") and a European female (labeled "EurF012H"). These last two skulls are included as they were taken to represent the most deeply divergent of over 200 modern skulls considered in preliminary analyses (results not shown). They were effectively outliers, not grouping with most members of their ethnic groups and are therefore at the limits of the diversity that can be expected in modern humans, being at roughly the 1/100 and 1/200 limits for such diversity, or the $99^{th}$ and $99.5^{th}$ percentiles of modern human variation.

The fossil skulls may be grouped into a number of sets. Sex assignments follow Harvati et al. (2011, 2013). Also indicated after the label are the rough cranial capacity, where known, of each skull, rounded to the nearest 100 cubic centimeters (cc). Thus, a size of 14 indicates an estimated volume of about 1400 cc. As a rough guide, one sample of modern humans put the modern male average at around 1440 cc, and females at around 1240 cc (Henneberg 1988). However, the later figure for modern females as been argued to be more realistically around 1330 cc (Rushton 1992). Thus a skull described as CroMag1_M17, for example, has the M after underscore indicating a male (a female is F and undetermined is U) with an estimated cranial capacity of about 1700 cc.

There is a good sized sample of skulls associated with the Gravettian culture, a European culture that appeared in Europe ~35kya and persisted until about 20 kya. Gravettian cultures were preceded by pre-Gravettian cultures that mark the earliest modern humans yet found in Europe, with modern dates extending back towards 45 kya (e.g., Kranioti et al. 2011). The following skulls are associated with Gravettian culture Cro Magnon (CroMag1_M17, CroMag2_F, CroMag3_U16), Dolni Vestonice (Dveston3_F13, Dvesto13_M15, Dvesto15_M14, Dvesto16_M15), Predmosti (Predmost3_M16, Predmost4_F13), Grimaldi (1955-skull4, Grim1955_U), Brno1_U16, and Pavlov1_M15, while those associated with Pre Gravettian culture are Mladec1_F15, Mladec5_M17, Pestera Oase (Oase2_F), Cioclovina (Cioclov_U15), and Muierii1_F). Of the more recent fossil skulls OhaloII_U from Israel is not assigned to either of these cultures, while the later Gravettian and post-Gravettian skulls of Abri Pataud (Abri_F14) and Chancelade (Chance_U15) are left unassociated later in constrained tree analyses. Endocranial volumes come from Holloway et al. (2004), except for the more recent measurement of Cioclovina (Kranioti et al. 2011). Cultural associations are indicated in Schwartz and Tattersall (2002, 2003, 2005) and Trinkaus (2007), for example.

Samples of early near anatomically modern humans that might represent hybrids with earlier archaics are the Qafzeh (Qafzeh6_M16 and Qafzeh9_F15) and Skhul (Skhul5_M15) specimens from Israel/Palestine and the Zhoukoudian Upper Cave Skulls (UC1_M15 and UC3_F13). Not anatomically modern humans, but commonly accepted as nearer to modern humans than Neanderthals are the African specimens from Singa (Singa_M16), Jebel Irhoud (Ihrhoud1_U13 and Ihrhoud2_U14), Iwo Eleru skull (IwoEleru_U) and Laetoli Ngaloba skull (LH18_U14). This last skull is sometimes classified as *Homo heidelbergensis* (Magori and Day 1983), but most researchers are currently associating it somewhere along the lineage leading to modern humans exclusive of Neanderthals (e.g., Schwartz and Tattersall 2005).

Neanderthals are represented by the skulls Guattari_M16, Chapelle_M16, Ferass1_M16 (Ferrassie), Quina5_F12, Feldh1_M15 (Feldhofer), Spy1_F13, Spy2_M16, Amud1_M17, Shanid1_M16, and TabunC1_F13. The last three are from the Middle East, while the remainder are from Western Europe. The poorly defined species *Homo heidelbergensis* often has included in it the European skulls Petralona (Petral_M12) and Sima de Los Huesos (SH5_M11), the Chinese Dali_U11, and the African Kabwe_M13 (Broken Hill) and Saldahna_U12 (Elandsfontein). Finally, there are the two early African *Homo erectus/ergaster* skulls ER3733_U8 and ER3883_U8. Nearly all these skulls are described in detail in Schwartz and Tattersall (2002, 2003, 2005).

The above assignments to various groups has particular importance later when constrained tree analyses are considered (Swofford et al. 1996) and in assessing whether a particular trend in part of the whole of the tree might be simple allometry, that is, shape changes associate with larger or smaller skulls. Further, the assignments to male and female of members of what are considered well founded populations or clades also allows an assessment of this factor to the form of the tree.

**3.1 Subtree extraction**

Adding in extra specimens to a distance based phylogenetic analysis can see the g%SD increase. This is analogous to the CI tending to increase as taxa are added to a parsimony-based analysis (Swofford et al. 1996), with the difference that the g%SD (adjusted for the degrees of freedom) should be stable. That is, if the extra data match these already in the analysis, including the proportion of external edge lengths they add to the tree (Waddell 2014). Here this last condition should hold approximately true, yet the g%SD rises from 6.16% for 13 taxa to around 7.68% as seen later. These fit values are for the optimal OLS+ tree, with the Procrustes distance transform set to the power $q = 0.5$. Before going deeply into the analysis of the much larger set of fossils, it is interesting to ask how this large amount of extra data may alter the conclusions reached on the smaller sets of data analyzed in Waddell (2014).

Comparison of the impact of the much larger set of data can be made fairly reliably via the method of subtree extraction applied to the residual resampling analyses (Waddell et al. 2011a). Rather than using the 13 taxa data set, the 12 taxon data set created by the removal of Qafzeh 6 may be a more powerful comparison. This is because this data set had a g%SD of only 4.34%, a considerable improvement from 6.16% with Qafzeh 6 included, and a bit more than half that of the 52 taxon data set considered here (7.68%). Note that in this instance the mean square error of the larger set to the extracted set differs by a factor of nearly four, so adjustment for this is considered.

The results are shown in figures 1 and 2. The tree in figure 1 for 12 taxa is that which is represented in Waddell (2014). These results, where the variance for residual resampling was set to four times that estimated from that subset of the data, may be compared to those in figure 8a of Waddell (2014). The following figure 2 is the subtree extracted from 10,000 independent residual resampling replicate data sets based on the variance of the 51 taxon data set and otherwise matched to the results in figure 1.

Comparing figure 1 and figure 2 the first thing to note is that the subtree extraction resampling shows two changes to the consensus tree. The first is that the Khoisan individuals form a moderately stable direct sister relationship in 80% of replicates, while LH18 and IwoEleru also have a sister group relationship, but in a more modest 59% of replicates. The first of these rearrangements suggests increased accuracy with the addition of more data, while the second is unclear. Otherwise, all other internal edges in the tree have increased support in the subtree extraction consensus tree compared to figure 1. Compared to figure 8a with its variance of a little over one fourth that of the new 51 taxon dataset, support values are still fairly similar. The decreases have been for the internal edge separating Iwo Eleru from LH18, which cannot yet be known as an improvement or not, while that involving the Khoisan seems to be both an improvement in accuracy and precision. The only other real differences in precision are moderate decreases on the three edges from 100% down to the high to mid 90's.

This suggests that the addition of this extra data may have significantly improved the accuracy of the subtree overall while also improving it precision compared to matching variance or about the same when variances are unmatched.. This is despite the fact that many specimens with a more conflicted signal about where they should fit in the tree, e.g. Qafzeh 6, have raised the overall residual errors. Perhaps the most interesting result is the reemergence of the possibility that the Iwo Eleru skull is an early divergence on the lineage leading to LH18, seen previously in NJ and BME analyses (Waddell et al. 2014).

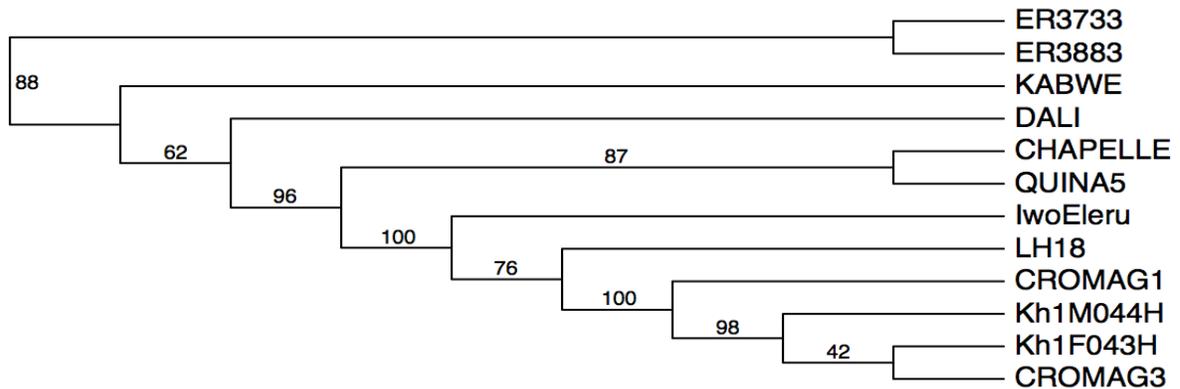

Figure 1. Residual resampling results for the 12 taxon dataset analyzed in Waddell (2014). The distances are used here were obtained by raising to the power $q = 0.5$ the sum of squared pairwise Procrustes coordinate differences, and tree selection was via OLS+ (Waddell et al. 2014). The variance used for resampling has been set to four times that of the original data, that is, mean g%SD of the residual resampling replicates is ~ 8.4. Since these are based on 10,000 replicates, the standard deviation of these counts due to sampling error is ~0.5 at 50 and ~0.3 at 90; thus, only the last digit is expected to fluctuate due to sampling error alone.

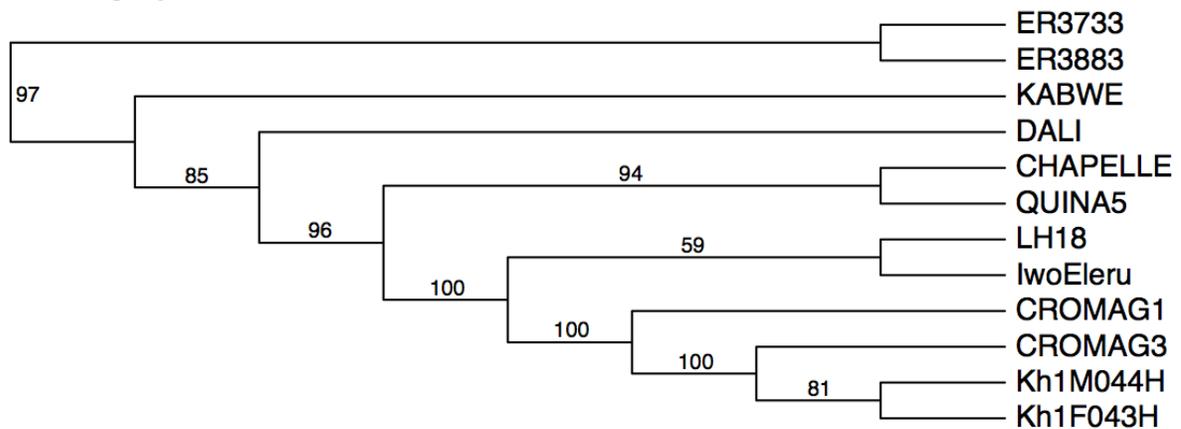

Figure 2. Subtree extracted from the data set of all fossil skulls and four recent moderns. Here the g%SD of the original data was 7.68% and $q$ was again 0.5 and tree selection was again via OLS+. Since this is based on 10,000 replicates, the standard deviation of these counts due to sampling error is again expected to be ~0.5 at 50 and ~0.3 at 90.

### 3.2 The tree of 51 skulls at $P = 0$ (OLS+) and $q = 0.5$

The first tree considered is based upon the transformed distances which maximized the likelihood of the 12 taxon dataset in Waddell (2014). This was at a power, $q$, of close to 0.5. In addition, $P = 0$, was used along with a prior on all edges having to be non-negative in length, which corresponds to non-negative ordinary least squares, abbreviated OLS+.

The resulting tree is shown in figure 3, while the support values based on 1000 residual resampling replicates is shown in figure 4. The form of the tree fits closely with general expectations from the literature including Schwartz et al. (2005). The two *Homo erectus* skulls cluster at the root, followed by Kabwe as the next earliest diverging. The next part of the tree is a little confused, largely because the Neanderthal skulls fail to form a clade as might be expected. Many of these interior edges are relatively weakly supported as seen in figure 4. A cherry of a male and a female Neanderthal branches off after Kabwe, then the Dali skull branches. The next branch contains the interesting cherry of Petralona with SH5 (the European specimens usually assigned to *Homo heidelbergensis*), with a strong support value of 99% (figure 4). Following this are more Neanderthals, then a strongly supported edge (100%) that marks the beginning of what, without quantitative evidence from evolutionary models, many paleoanthropologists have suspected is the African lineage out of which arises modern humans.

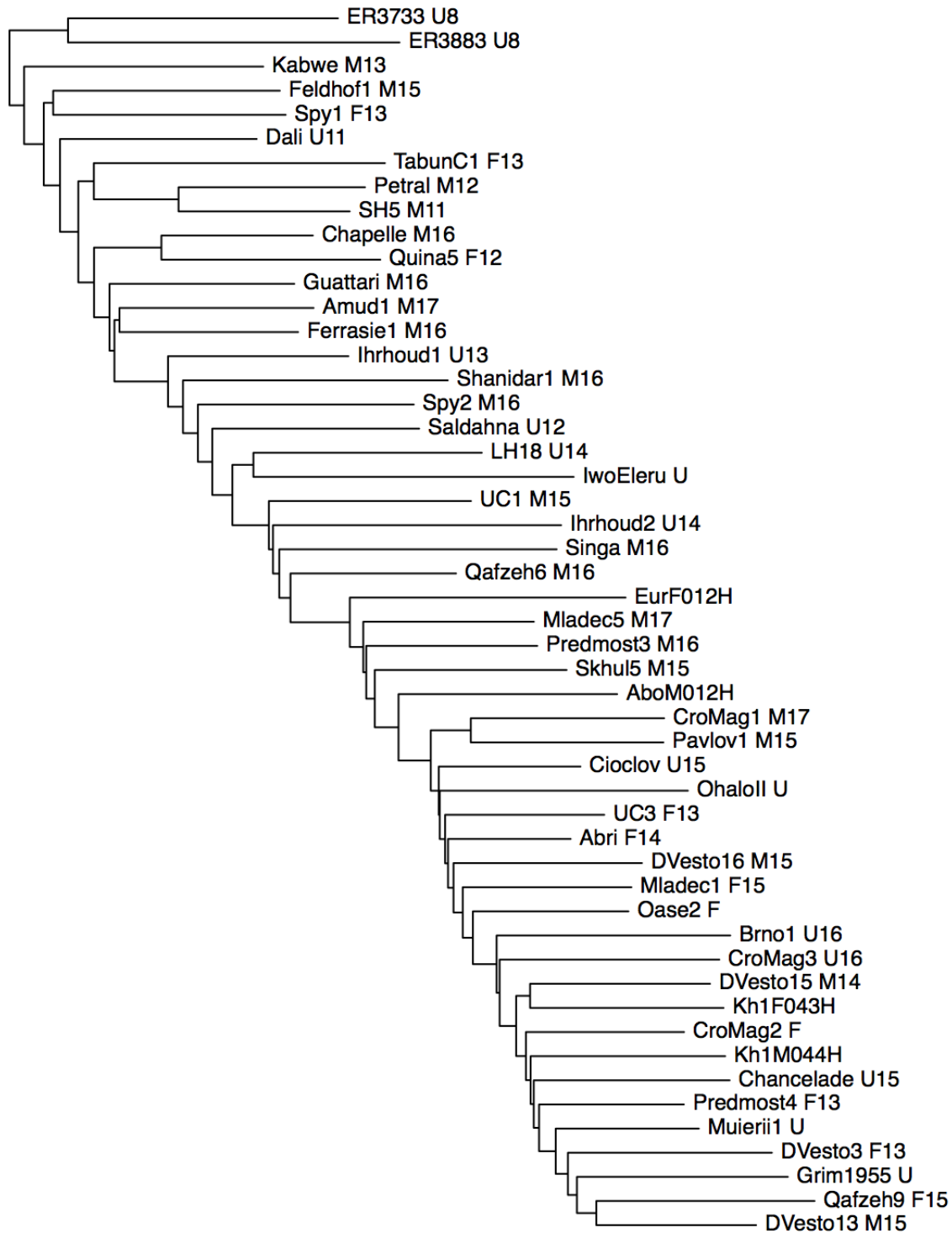

Figure 3. OLS+ tree based on $q = 0.5$ power transform of the Procrustes sum of squared coordinate differences. The g%SD is 7.678 with k = 99. After each skulls label is the indicator of the assigned sex and then the cranial capacity rounded to the nearest multiple of 100 cubic centimeters (cc).

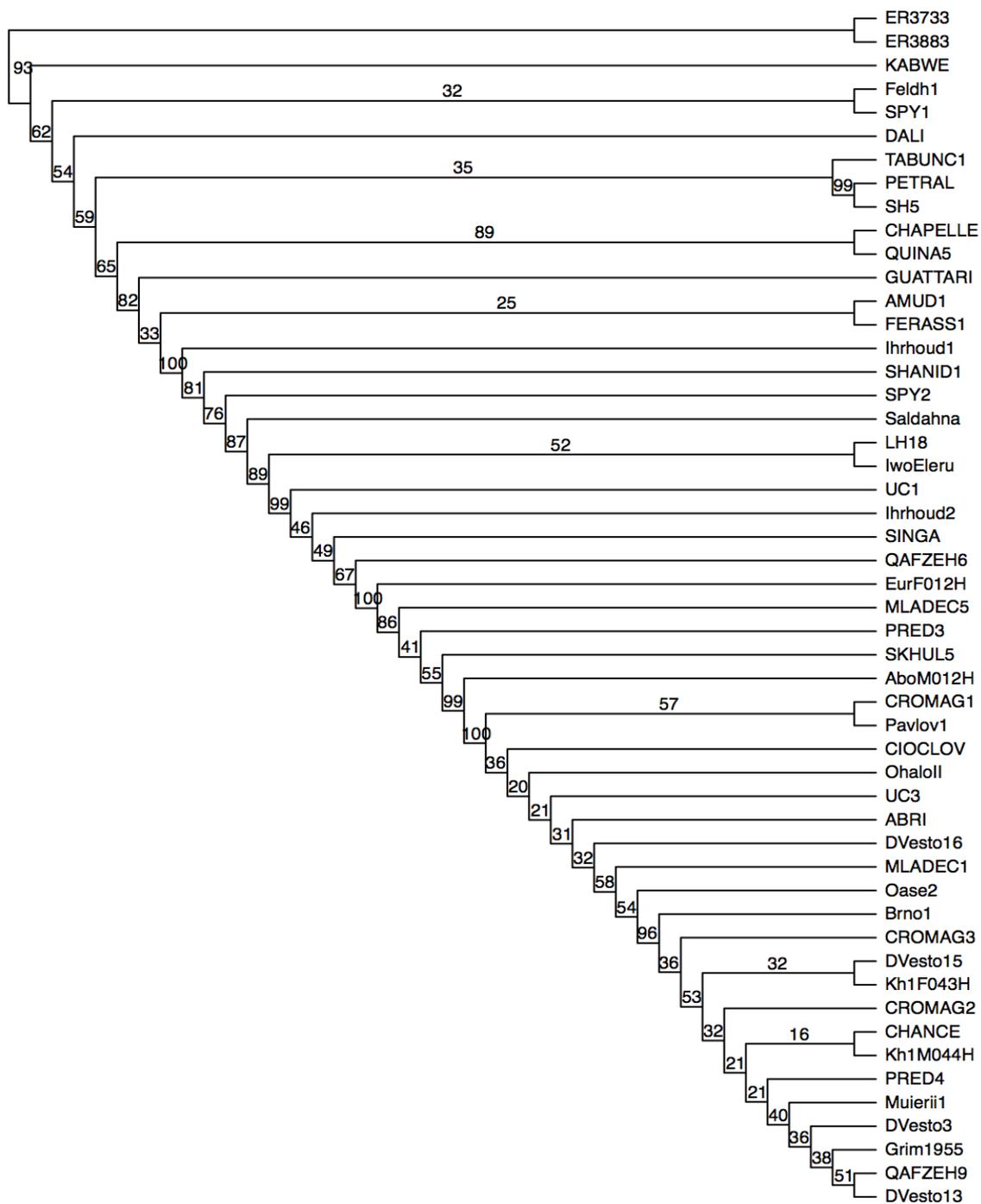

Figure 4. Consensus tree of 1000 residual resamplings of the OLS+ tree based on $q = 0.5$ power transform of the Procrustes pairwise sum of squared coordinate differences.

Looking at the labels for sex and cranial size as shown in figure 3, depth in this tree does not seem to be any clear consistent association with either property. For example, large and smaller brained Neanderthals occupy different depths as do males and females. Near modern humans at the bottom of the tree there is also not a hugely strong association of position, except that there is a predominance of skulls assigned female or undetermined and of a smaller average cranial capacity nearest to the modern Khoisan skulls. More strikingly, within the UC (UC1), Qafzeh (6), Gravettian (Predmost3, CroMag1 and Pavlov1) and pre-Gravettian assemblages (Mladec5), the deepest members are always males with a clear tendency to have larger than average cranial capacities. Thus there does not seem to be any simple evidence for allometry determining the overall structure of the tree. However, the unusual pattern to the placement of large headed males in probably hybrid populations will be explored further below.

The lineage after the Neanderthals leading to modern humans seen in figure 3 is well defined with a support value of 100% (figure 4) but there are two notable caveats to its anticipated composition. These are the two Neanderthal skulls Shanidar 1 and Spy 2. For some reason these are being drawn down the African lineage towards modern humans. It is not known why, but is perhaps useful to note that the Spy Neanderthals are some of the very latest known, and it is not impossible that there has been some modern human introgression with them.

The major argument against Spy2 being a hybrid with modern humans is the suspicion that in Europe, this was a fast moving and fully effective invasion of moderns that left little time for such a hybrid population to exist (that is hybrid population of mostly Neanderthal). However, that argument is weaker in the case of Shanidar 1, a late Middle Eastern Neanderthal in a region that seems to have had potential hybrid boundary of anatomically or near anatomically modern humans and Neanderthals for tens of thousands of years. Where exactly that boundary was at different times, or how broad it was, is unclear (Bar-Yosef and Belfer-Cohen 2011). However, the fact that the Qafzeh 6 skull shape looks very much like a hybrid (Waddell 2014) suggests it was potentially a fairly broad and long lived boundary, else why would one in three of near anatomically modern forms at Qafzeh and Skull seem to show this so clearly? Alternatively, and not necessarily exclusively, Shanidar and Spy2 are just outliers of the scatter of Neanderthal skull forms on this tree. That is, the data are noisy, and the exact position of a single skull on the tree must be suspect, but the relative position of samples of multiple individuals of a well defined population should be clearer, which is something to be explored later using constraint trees.

On this African lineage leading to modern humans, Irhoud 1 stands out at a particularly early form. It is separated from Irhoud 2 by a few well supported internal edges, yet Schwartz and Tattersall (2002, 2004) do not suggest different morphs at this location. It is particularly interesting that the Saldahna skull also appears as an early member of this lineage. It will be important to see what robustness this finding has as parameters of the phylogenetic analysis are changed. Following Saldahna there are other middle to late Pleistocene African forms represented by Singa, Irhoud 2, and LH 18 (quite probably in the range of 150 to 300 thousand years old, Millard 2008). Again, Schwartz and Tattersall (2004\5) give probably the best assessments to date of their affinities based on qualitative traits, and there are only weak hints of what their relative positions might be based on these traits. The one surprise in here is UC1, a male from China, perhaps only about 20,000 years old. It shows up surprisingly deep and it also displaced far from the other specimen from this site, UC3, which falls much closer to typical modern humans. This must raise suspicion that this too is a hybrid of some sort (archaic lineage unknown). Following Qafzeh 6 in the tree, there is another strongly supported internal edge (figure 4).

Broadly speaking, the last part of the tree may be partitioned into forms that might be called close to typical modern forms, marked by the positions of the Khoisan individuals, and a broad range of forms which are not typical modern humans. The modern European female and the modern Australian Aboriginal skull mark approximately the $99^{th}$ quantile of modern variation. Note that many, but not all, of the early European skulls fall in this region, while Mladec 5, which Trinkaus (2007) notes shows multiple Neanderthal traits, falls very deep.

There is the interesting pattern for both the Gravettian and the pre-Gravettian skulls that the female skulls tend to fall very close to, or even within, the diversity marked out by the two Khoisan skulls. So too do the undetermined sex skulls of these two groups, both in contrast to the males, which can fall much deeper. One possibility here is that since sex determination for modern humans can rely upon "male" features which are also potentially confounded with the features of modern/archaic hybrids (e.g. brow development), there is a bias towards calling individuals with more archaic effects on the skull, male. This directional bias of males appearing more archaics does not occur within Neanderthals (figure 6), nor is it so prevalent with most populations of modern skulls (results not shown).

Of the pre-Gravettian individuals, their order of "archaicness" based on this tree would seem to be Mladec 5, Cioclovina, Mladec 1, then Oase 2. Clearly the Muierii skull is closest to a typical modern being tightly grouped with fully modern forms, and the order is nearly exactly opposite that predicted in the introduction. This would seem to suggest that the archaic features/genetic effects emerging in this analysis of general skull shape do not have the same genetic basis as the skull shape features of these specimens pointed out in Trinkaus (2007).

### 3.3 OLS+ with $q = 0.5$ and constrained evolutionary trees

The purpose of this section is to further explore the relationships of these skulls with the aid of constraint trees. The first set of constraints is forcing populations and/or morphs well defined elsewhere to be monophyletic. Except for Neanderthals these are well prescribed fossil locations where Tattersall and Schwartz (2002, 2003, 2005) do not raise significant concerns about more than a single morph being present. It is for this reason that Qafzeh specimens are not grouped, since Schwartz and Tattersall (2003) note the diversity of forms (morphs) at this site.

Figure 5 shows the result of fitting these constraints and then assessing the robustness of the tree with 1000 residual resamplings. Again, starting at the root, note that all heidelbergensisy things other than Saldahna predate the Neanderthal split with 100% support. They do so, however, without monophyly, but only moderate support for the exact branching order, except that the association of SH5 and Petralona is again strongly supported. Within Neanderthals there are no really strong associations of particular individuals despite being constrained into a single clade.

Figure 5 shows there is now 100% support for the post Neanderthal split African lineage that includes modern humans. On this lineage, Saldahna appears as the first branch, with the Irhoud specimens second, and then a moderately supported LH18/IwoEleru clade. The relative order of appearance of these lineages has about 80% support at each step. Following these there is another strongly supported edge (96%), then Singa being moderately deeper than Qafzeh 6. There then another strongly supported internal edge. Next, the unusual European Female skull (012) defines an ~ 99.5% quantile for the type of diversity expected within modern humans. The Zhoukoudian Upper Cave sample from China has unusually archaic skull shapes, and by this analysis, substantially more so than the Mladec individuals from Early Europe. This in turn suggests, based upon limited evidence, that the extent of interbreeding between near anatomically modern humans and archaics was at a higher frequency in this region. This would seem to fit in with reports of very unusual morphologies that persisted surprisingly late in China (e.g. the Red Deer people, Curnoe et al. 2012), that will also need to be more carefully assessed as a potential hybrid population. Complicating the hybrid interpretation in China is the fact that neither the near modern forms to be found in China, nor the late archaics they would be expected to have encountered are well defined (Jinniushan and Dali are possible examples, Schwartz and Tattersall 2003).

Considering now the very bottom of the tree, with typical early modern human forms defined by the two Khoisan skulls, it is notable that quite a few fossil skulls fall right down in this region. These include Qafzeh 9, the first known anatomically modern human. Some whole Gravettian samples also appear in this region, for example the material included from Dolni Vestonice. The Gravettian Cro Magnon and Predmosti samples appear a bit more archaic, that is, earlier diverging. The later European skulls of Abri Pataud and Chancelade, which still predate a major genetic turnover in Europe with the arrival of agriculture, also show evidence of somewhat archaic shapes.

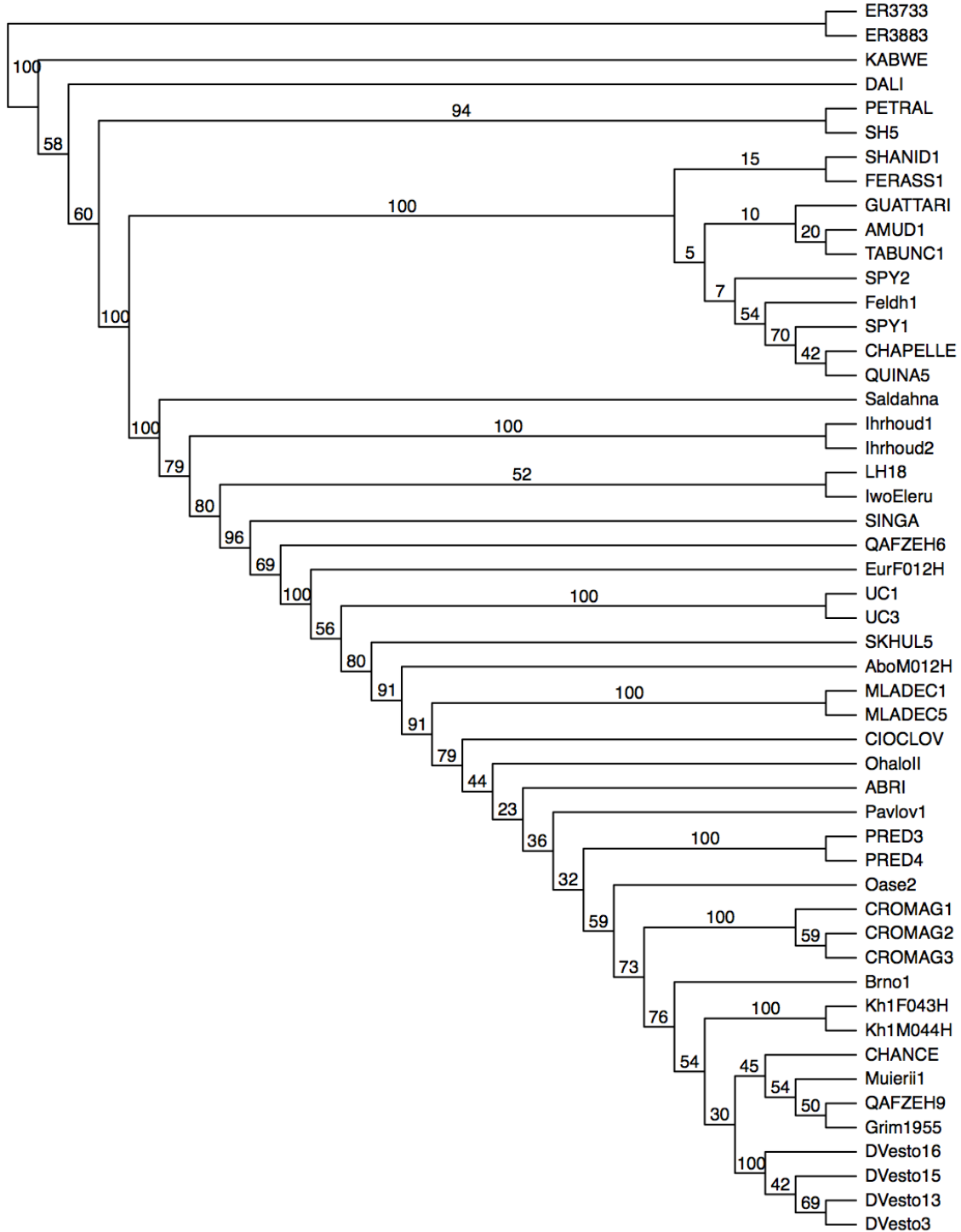

Figure 5. Consensus tree of 1000 residual resamplings of OLS+ tree based on $q = 0.5$ power transform of the Procrustes sum of squared coordinate differences and using the first constraint tree of readily defined populations/morphs. Except for Neanderthals these are well prescribed fossil locations where Tattersall and Schwartz do not raise concerns about more than a single morph being present. It is for this reason that Qafzeh specimens are not grouped.

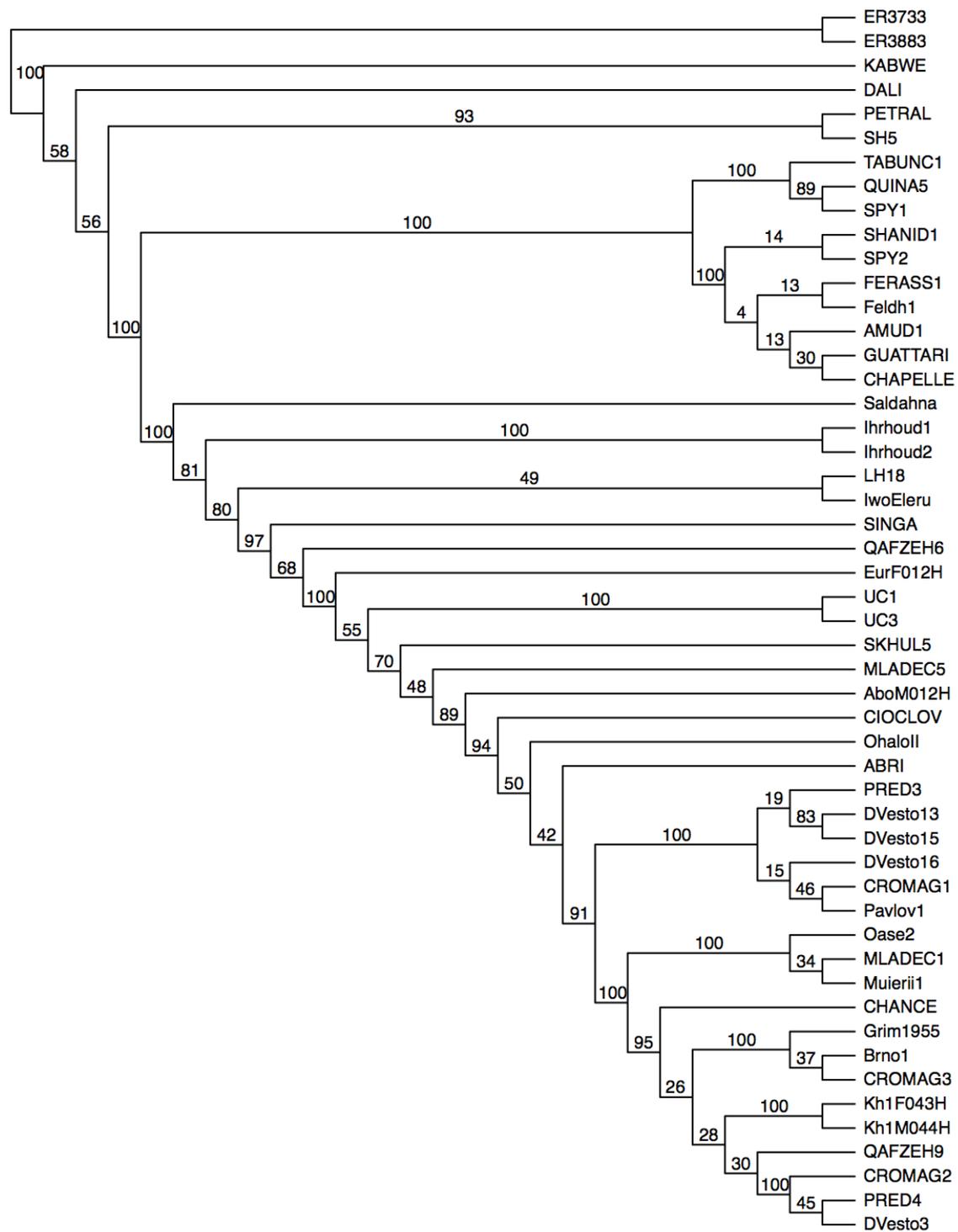

Figure 6. Consensus tree of 1000 residual resamplings of OLS+ tree based on $q = 0.5$ power transform of the Procrustes sum of squared coordinate differences and a constraint on populations and sex determination. The populations/groups used in figure 5 are with the additional grouping of Gravettian and pre-Gravettian and then a split into males, females and undetermined for these two groups plus Neanderthals.

Figure 6 considers a different constraint tree. In this example males, females and undetermined sex of the Gravettian, pre-Gravettian and Neanderthal samples are grouped in order to look further at possible biases in their phylogenetic position due to sex. For the Neanderthals, splitting into males and females surprisingly sees the female group with 100% support branch before the male group (not shown). That is, the female Neanderthal upper skull form appears more archaic on average compared to the males of the species. In contrast, for the Gravettian and pre-Gravettian samples there is a strong bias towards the female skulls appearing most modern, then the undetermined and most archaic are the skulls sexed as male. This result reinforces the possibility that within the Gravettian and pre-Gravettian samples, some archaic features, perhaps including shape and size, are somehow being confounded with primary and secondary sexual characteristics. Figure 6 also shows conclusively that the Gravettian and pre-Gravettian males are, on average and not just the most extreme examples, joining the tree deeper than the females. There are multiple internal edges with near 100% giving clear support to this interpretation.

**3.4 Optimal values for the Procrustes transform $q$ and its impact upon the tree**

It is useful to look at how the tree might be affected by power transforms of the sum of squared pairwise Procrustes coordinates between skulls, following the methodology and nomenclature of Waddell (2014). As figure 7 shows, the optimal value of $q$ with OLS+ is close to 0.26, with a well defined single optimum. This power transform compared to the power transform that was optimal for the 12 taxon data in Waddell (2014) is approximately proportional to the transform encountered going from neutral genetic theory to the 12 taxon data encountered in Waddell (2014). It is quite a powerful transform and may be suggesting that most of the error in fit between the data and a tree is due to convergence of skull shape. This in turn can be consistent with hybridization of taxa, but also limited possibilities for major trajectories of skull shape change.

The OLS+ tree with $q = 0.26$ is shown in figure 8. It is very similar to the tree at transform $q = 0.5$ shown in figure 3. There are a very few local rearrangements such as Irhoud 2 and Singa now being sister to each other. As seen in Waddell (2014), the optimal $P$ value is correlated with the $q$ value (results not shown). For this data, the optimal $P = -0.043$ value at $q = 0.5$ is very close to 0.

When allowing $P$ to be optimized with $q$, the optimal value for $q$ gets very small. Below 0.05 there were some computational issues, but it appears that a plateau of likelihood has been encountered. At $q = 0.05$ the optimal value for $P$ used for the weights of WLS+ is -19.77 . The resultant WLS+ tree shown in figure 9, has a few interesting fairly local rearrangements. Most of the Neanderthal lineages now form a clade with all the Eurasian *Homo heidelbergensis* specimens. Within this group, the Eurasian *Homo heidelbergensis*-like specimens are also monophyletic. Notably, Spy2 and Shanidar 1 remain in positions drawn towards modern humans. Another group to appear is UC3, Abri Pataud and Cioclovina, which are classified as modern female forms but from quite different regions and times.

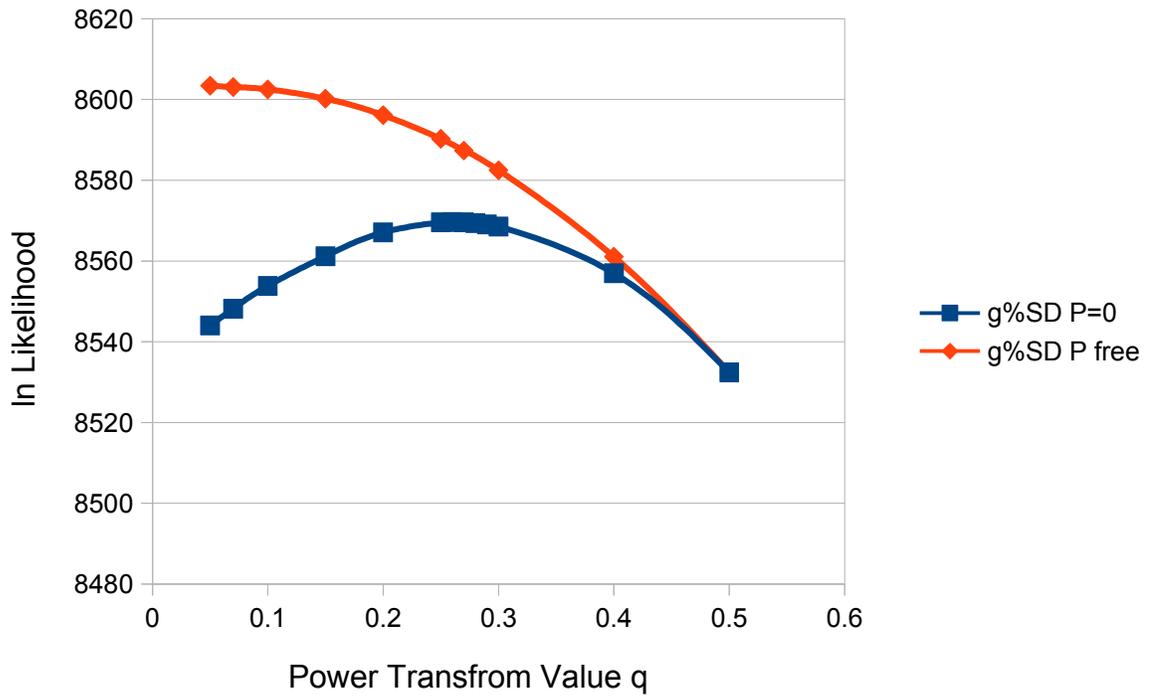

Figure 7. A plot of the natural log likelihood of the data versus the power transformation q applied to the sum of squared pairwise Procrustes coordinates. Values of q below 0.05 were not plotted as computational issues were encountered, perhaps caused by optimal weight power values of -100 and greater. The optimum for OLS+ is ~0.26, while the optimal value for WLS+ with P free here is 0.05.

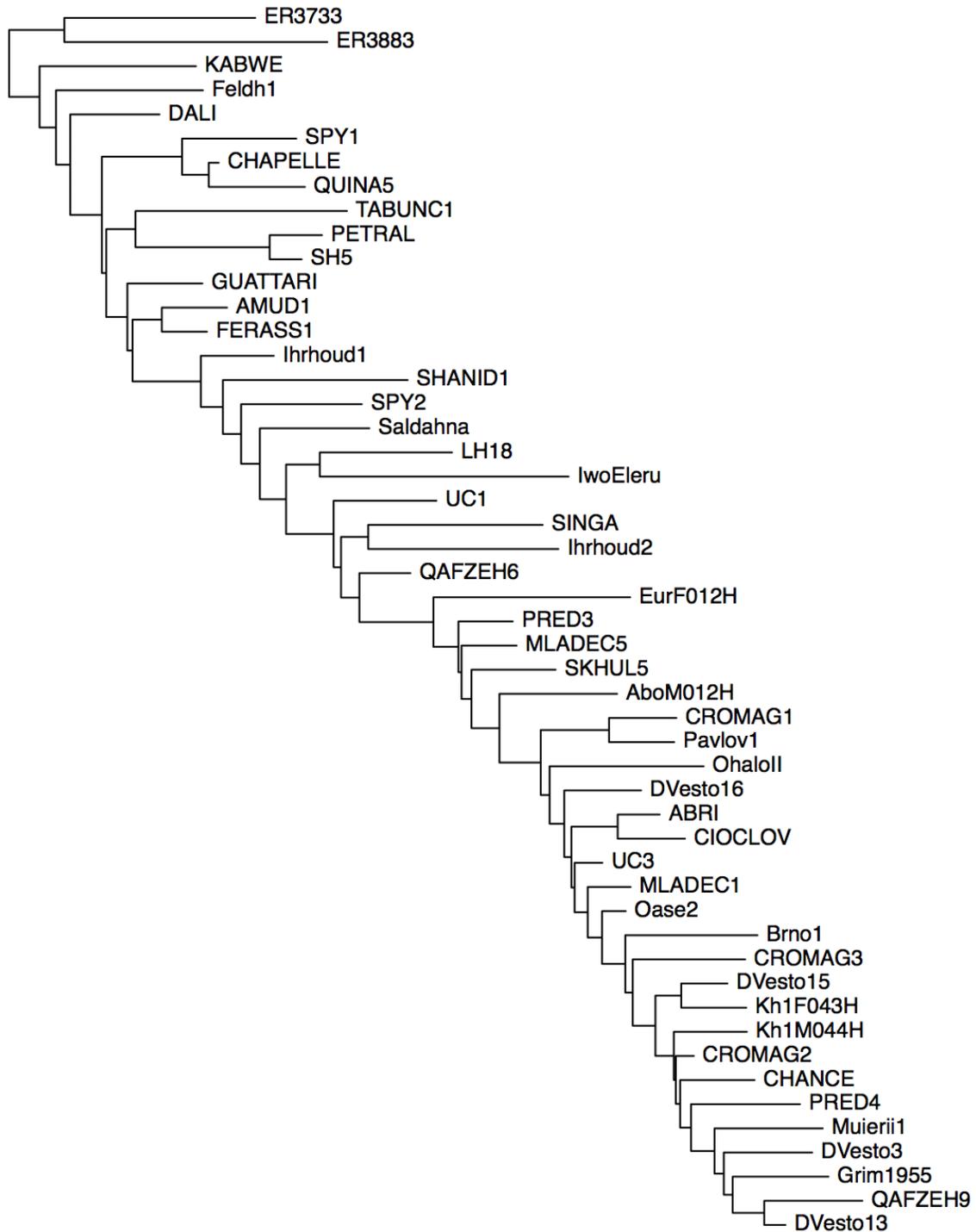

Figure 8. The OLS+ tree with power transform $q$ = 0.26 applied to the sum of squared Procrustes coordinates. External edge lengths have been shortened by 0.1 to make the internal edges more obvious.

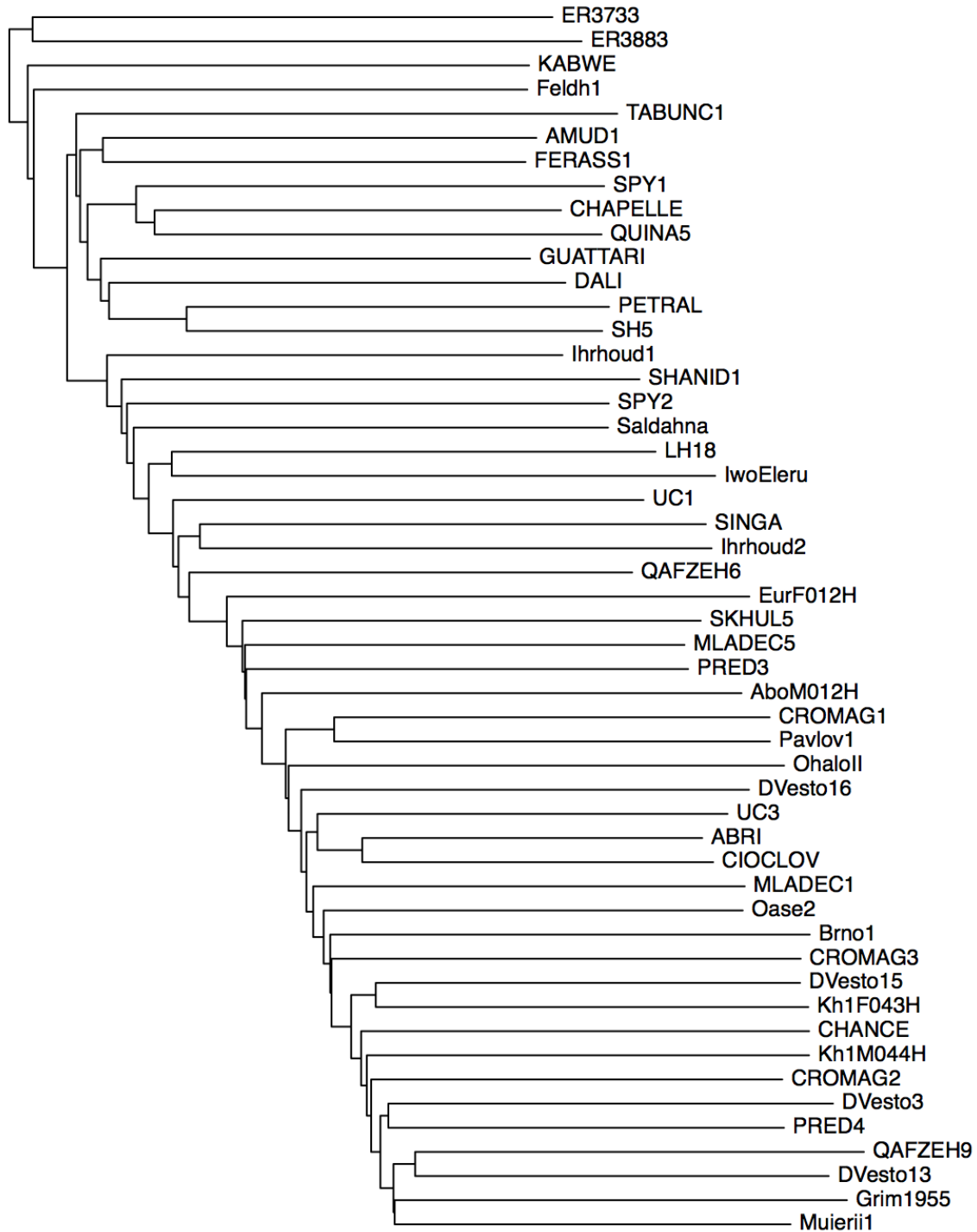

Figure 9. The WLS+ tree with optimal $P = -19.77$ and power transform $q = 0.05$ applied to the sum of squared Procrustes coordinates. External edge lengths have been clipped 0.3 to make the internal edges more obvious.

# 4 Discussion

The current analyses provide strong support for the results of Waddell et al. (2014). For example, the enigmatic Iwo Eleru skull is seen to with 100% confidence to locate in a part of the tree well outside the range of modern human variation, and with one exception (UC1), amongst a range of middle Pleistocene African archaic forms. One exception to reinforcing the results in Waddell (2014) is a local rearrangement of the position of Iwo Eleru. However, while IwoEleru now locates most often sister to LH18, the hypothesis that it represents a preceding lineage cannot be excluded.

Schwartz and Tattersall (2005) give a very useful summary of the morphological forms (morphs) at different sites and times. For the African forms at about 2 to 1.5 mya they describe multiple distinct forms which cross the traditional *Homo habillis/Homo erectus/Homo ergaster* lines. Represented here are the 3733 and a 3883/3732 morphs, which are described as quite distinct on discrete morphological features. Here, the shape of the major sutures and boundaries of the skull cap are compared. The difference between these two morphs in terms of shape of the skull cap is quite comparable in magnitude to the difference seen between two Neanderthals or between two modern Khoisan, for example. In this sense they do not strongly contrast. However, in terms of shape distance-based phylogenetic analyses, this is not that unusual, in that much of the material of this period and often lumped (Lordkipanidze et al. 2013) under the title *Homo erectus* from all around the world has broadly similar shape, despite some appreciable size differences (e.g. brain size varying by a factor of two), as well as huge geographical and temporal differences. To this, Schwartz and Tattersall (2005) would argue also major differences in discrete qualitative morphological features. Indeed there is quite a controversy based largely on interpretations of non-evolutionary shape analyses (e.g., Schwartz et al. 2013, Spoor 2013, Bermúdez de Castro et al. 2014, Waddell 2014). Taken together, this suggests some conservatism of basic skull cap shape for this diverse assemblage, and a correspondingly worse signal to noise ratio in the data. This conservatism in overall skull cap shape change accelerated with the middle Pleistocene and accelerated again with the African lineage leading to *Homo sapiens* (Waddell 2014). It may also explain why non-evolutionary assessments of 3D geometric morphometric upper skull shape data are failing to identify distinct assemblages in a huge range of *Homo* specimens.

The sister group relationship of Petralona and SH5 is surprisingly strong in the analyses presented here, and is not predicted by the lists of morphological features considered by Schwartz and Tattersall (2005). However, they do conclude that what they consider an endemic European lineage of SH5, Neanderthals and allies is probably linked to some of the forms commonly called *Homo heidelbergensis*. Perhaps is is no coincidence that of all the *Homo heidelbergensis*-like things in this analysis, SH5 and Petralona, the only two European forms, are bracketed by Neanderthals, and at the more extreme values of $q$ with $P$ also adjusted, they form a clade with Dali, the other Eurasian "*Homo heidelbergensis*" within a large group of Neanderthals. This is on marked contrast to the two most prominently cited proposed African *Homo heidelbergensis* skull caps, Kabwe and Saldahna, which consistently fall well away from Neanderthals. Indeed, the positions of these two taxa on the tree is in contrast both with the Eurasian *Homo heidelbergensis* specimens and between themselves. That is, always deepest diverging on the tree, for Kabwe, and always near the root of the African lineage persisting after the divergence of Neanderthals and heading to modern humans, for Saldahna. The residual resampling results supporting such distinct positions was nearly always 100%. This is all more evidence that *Homo heidelbergensis* is probably not a clade, but has members scattered on at least three and possibly four major lineages in the tree of *Homo* (the last being achieved with the data used here when LH18 is described as a member of *Homo heidelbergensis,* Magori and Day 1983).

In these all analyses, the first well defined anatomically modern human, Qafzeh 9, consistently falls within the diversity marked out by the male and female Khoisan skulls. Thus giving it the title of the "Skull of Eve" is less of stretch than calling the ancestor of our mitochondrial DNA

"Eve" (e.g., Penny et al. 1995, Waddell and Penny 1996). While Qafzeh9 is very modern in overall cranial shape, it does exhibit some characteristics that suggest it too may be part of a hybrid population (between modern or near modern humans and Middle Eastern Neanderthals). This includes a planum alveolare on the mandible, which is a pleisiomorphic trait which Trinkaus (2007) sees as introgressed from Neanderthals into early europeans also. Importantly, the fully modern shape of the upper skull of Qafzeh 9, as well as of many early Europeans, would seem to convincingly deflate the often heard argument that earlier modern humans only look more archaic due to life style or climate (at least as regards upper skull shape). It seems the favored hypothesis now must be that if a skull looks decidedly archaic, then archaic genetic factors are the leading explanation.

These phylogenetic analyses also reveal a trend for large brained males in probable hybrid populations including pre-Gravettians, Gravettians, the Upper Cave and Qafzeh/Skhul specimens to show markedly more archaic skull shapes. This may simply be a sex assignment bias, but more interestingly, it might also be due to how archaic X chromosomes in a hybrid context are imprinted and guide brain development (e.g., Lepage et al. 2013). This trend is not seen in Neanderthals or in recent modern human populations, making its biological reality seem more likely. The overall result of these analyses is to suggest that the extent of archaic genes was Gravettian, pre-Gravettian, Qafzeh/Skhul and Upper Cave China, in order of increasing archaic content.

There has been much discussion of how much and what archaic genes in modern humans means for the biology *Homo sapiens* (e.g. Krause et al. 2010, Reich et al. 2010, Waddell et al. 2011, Waddell 2013 Fu et al. 2015). To date analyses only reveal a few percent of archaic genes in the bulk of modern humans, with some small and localized populations such as Papuans and Australian Aborigines showing perhaps 10% of such genes across the genome from multiple sources (Reich et al. 2011, Waddell et al. 2011, 2013). However qualitative results such as those of Trinkaus (2007) and now the quantitative evolutionary analyses of this article show that early populations of near anatomically modern and even behaviorally fully modern humans appear to have incorporated a markedly higher percentage of archaic genes. Despite occupying vast areas such as Europe and China, most of this archaic genetic contribution appears to have died out, just as the archaics themselves did (or were died out). The ultimate reason for that remains unknown, although the mechanism in large areas such as Europe and China includes the wholesale replacement of earlier populations by agriculturalists. The question then becomes, why did these later highly successful populations not incorporate high percentages of archaic genes when the original modern occupants of the land had them in greater abundance? A leading hypothesis must be that the same coordinated gene complexes that made the ancestors of the crown group *Homo sapiens* such an irresistible force may well have again asserted themselves in the processes leading to the rise of the first really successful post hunter gather societies. How much of a role natural selection might have played in these contexts remains to be seen, yet the role for natural selection weeding out much of the legacy of archaic interbreeding cannot be dismissed.

## Acknowledgements

This work was partly supported by NIH grant 5R01LM008626 to PJW. Thanks to Katerina Harvati, Yunsung Kim, Hiro Kishino, Jeff Schwartz, Dave Swofford, and Xi Tan for helpful discussions and assistance with data, software and calculations.